\documentclass[aps,prl,twocolumn,nofootinbib]{revtex4-2}
\usepackage{orcidlink}
\usepackage{array}

\usepackage{graphicx}
\usepackage{dcolumn}
\usepackage{bm}
\usepackage{xcolor}
\newcommand{\purple}[1]{\textcolor{purple}{#1}}

\begin{document}


\title{Parity-violating optical rotation in the Th-229 nuclear clock transition}

\author{S. Lahs
\orcidlink{0009-0003-1041-5920}}
\email{sebastian.lahs@tuwien.ac.at}
\author{T. Schumm
\orcidlink{0000-0002-1066-202X}}
\affiliation{Atominstitut, TU Wien, 1020 Vienna, Austria}

\author{V.~V. Flambaum \orcidlink{0000-0001-8643-7374}}
\email{v.flambaum@unsw.edu.au} 
\affiliation{
 School of Physics, University of New South Wales, Sydney 2052, Australia
}

\author{D. DeMille
\orcidlink{0000-0001-7139-4121}}
\affiliation{Department of Physics and Astronomy, Johns Hopkins University, Baltimore, MD 21210, USA}

\date{\today}
\begin{abstract}
The exceptionally low-energy nuclear transition in $^{229}$Th provides a
new opportunity to study hadronic parity violation (PV) by optical methods. The
PV nucleon--nucleus interaction admixes opposite-parity nuclear
states into the $5/2^+$ ground state and the $3/2^+$ isomer, thereby inducing
an electric-dipole amplitude $E1_{\rm PV}$ that interferes with the magnetic-dipole
clock transition amplitude $M1$. In a single-valence-neutron
Nilsson model, we find $|E1_{\rm PV}/M1| \sim 2\times10^{-5}$. This interference can be detected as a parity-violating nuclear optical rotation (PVNOR)
of resonant laser light transmitted through a $^{229}$Th-doped crystal. For
a highly-doped CaF$_2$ crystal fiber, the estimated rotation is of order $10^{-6}\,{\rm rad}$. A crossed-polarizer
measurement can reach this sensitivity with existing ultraviolet laser technology. We also analyze systematic effects in the form of optical birefringence,
stray magnetic-field-induced Faraday rotation, and local crystal defects. Such a
measurement would provide the first direct probe of the PV neutron-nucleus
interaction constant, $g_n$.
\end{abstract}

\maketitle

\section{Introduction}
Studying the effect of the weak interaction in low-energy bound systems such as atoms and nuclei is a longstanding effort, complementary to experiments at high energy colliders~\cite{guena2005atomic, ramsey2006hadronic}. In particular, the parity-violating (PV) component of the nucleon--nucleon interaction is a
low-energy manifestation of the electroweak interaction between quarks. Its
experimental study is difficult because weak nuclear amplitudes are masked by
the much larger strong and electromagnetic interactions. Nevertheless,
PV observables have been measured in nucleon scattering, compound
nuclear resonances, nuclear electromagnetic transitions, and atomic systems
through nuclear anapole moments; see, for example,
Refs.~\cite{henley1969parity,FlaKhrSus84,FlaMur97,Mansour,Fadeev}. These
measurements are complementary because they probe different combinations of
the underlying PV meson--nucleon couplings and have very
different nuclear-structure uncertainties.

For single-particle estimates, the PV nucleon--nucleus
interaction is conventionally parameterized by effective proton and neutron
constants $g_p$ and $g_n$ \cite{FlaKhrSus84}. The nuclear anapole moment measured in
$^{133}$Cs, an odd-proton nucleus, primarily constrains the proton constant
$g_p$. By contrast, the low-lying states of the odd-neutron nucleus
$^{229}$Th are dominated by a valence neutron and therefore provide direct
sensitivity to $g_n$. In the meson-exchange description, $g_p$ is dominated
by the sum of the $\rho$-meson and pion contributions, whereas these
contributions enter $g_n$ with opposite signs. 
This
information is also complementary to compound-nucleus experiments, whose
large statistical and kinematical enhancements~\cite{SushkovFlambaum1980ZhETF,FLAMBAUM1984} come at the cost of an interpretation based on ensembles of resonances and statistical nuclear theory. Note that the compound-nucleus results are mainly sensitive to $g_p$, which is expected to be larger than $g_n$~\cite{Vorov}.

The $^{229}$Th nucleus offers a qualitatively new route to such a
measurement. Its $3/2^+$ isomer lies only about $8.4\,$eV above the $5/2^+$
ground state, placing a nuclear transition within the range of narrow-band
laser spectroscopy. The transition is predominantly magnetic dipole (M1)
and forms the basis for nuclear clocks~\cite{de2026thorium, huang2026nuclear, Wang2026solid}. Recent progress in the
spectroscopy of $^{229}$Th-doped crystals has established laser-accessible nuclear
excitation, narrow resonances, and appreciable resonant optical depth
\cite{ooi2026frequency, morawetz2026continuous}. A solid host can contain a macroscopic number of thorium
nuclei in a compact, optically accessible sample, making it attractive for
transmission measurements.

The parity-violating nuclear interaction admixes negative-parity states into
both clock states. Analogous to parity-violating optical rotation in atoms \cite{ 
BarkovZolotorev1978, 
birich1984nonconservation,
meekhof1993high,edwards1995precise},
this effect can be observed by measuring the optical rotation of linearly polarized light close to the $^{229}$Th transition.

\section{Estimation of the size of the effect}
\label{sec:nilsson-pv-estimate}
\begin{figure*}[t]
\centering
\includegraphics[width=1\textwidth]{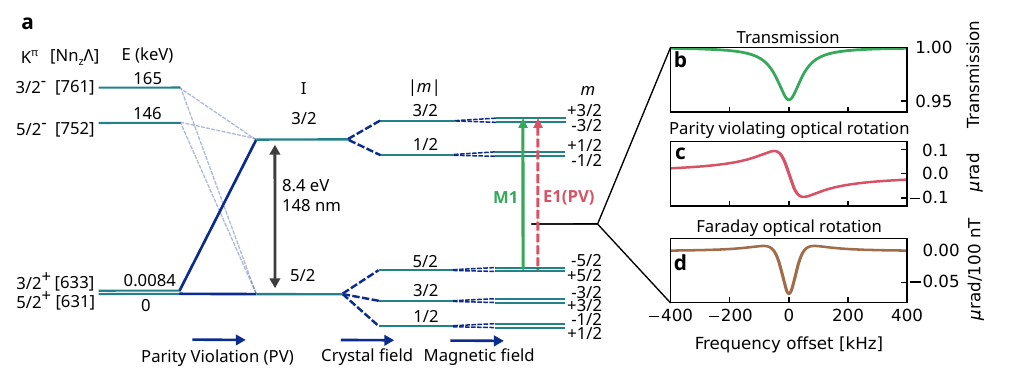}
\caption{\textbf{a}, PV nuclear interactions mix the ground and isomer states with higher-lying negative-parity states. In the crystal environment, the degeneracy of the $m$ sublevels is further lifted. \textbf{b}, Lorentzian transmission spectrum for the $m=\pm5/2$ to $m=\pm3/2$ transition. \textbf{c}, estimated magnitude of the PVNOR as a function of the frequency offset from line center. \textbf{d}, estimated Faraday optical rotation.}
\label{fig:energy}
\end{figure*}
We estimate the parity-violating (PV) electric-dipole amplitude in a
single-valence-neutron Nilsson model. The strong-coupling assignments of the
clock states are
\begin{eqnarray}
 |g\rangle=|I K^\pi=5/2,5/2^+;[633]\rangle,
 \qquad \\
 |m\rangle=|I K^\pi=3/2,3/2^+;[631]\rangle .
 \label{eq:pv-clock-configurations}
\end{eqnarray}
The closest experimentally established opposite-parity bandheads with the
same values of $K$ are $5/2^-[752]$ at $146.36\,\mathrm{keV}$ and
$3/2^-[761]$ at $164.53\,\mathrm{keV}$
\cite{pvThRuchowska2006,pvThLitvinova2009}. We use the conventional one-body
neutron weak potential
\begin{equation}
 H_W=\frac{G_F}{2\sqrt{2}\,m_N}g_n
    \{\bm{\sigma}\!\cdot\!\mathbf p,\rho(\mathbf r)\}
 \simeq
 \frac{G_F g_n\rho_0}{\sqrt{2}\,m_N}
    \bm{\sigma}\!\cdot\!\mathbf p,
 \label{eq:pv-weak-potential}
\end{equation}
where
$\rho_0=3/(4\pi r_0^3)=0.138\,\mathrm{fm}^{-3}$ for
$r_0=1.2\,\mathrm{fm}$. 

The constants $g_p$ and $g_n$ characterize the parity-violating
nucleon--nucleus potential. Following
Refs.~\cite{FlaKhrSus84,FlaMur97,Mansour,Fadeev}, for single-particle
estimates one may use the values
\begin{equation}
 g_p=3.4,\qquad g_n=0.9,
 \label{eq:gpgn_app}
\end{equation}
obtained using the meson-exchange model. These numerical values of $g_p$ and
$g_n$ correspond to the central values of the 
experimentally determined 
ranges of the
meson--nucleon interaction constants~\cite{Mansour,Fadeev}. The proton
constant $g_p$ is dominated by the sum of the $\rho$-meson and pion
contributions, whereas for the neutron the corresponding contributions enter
as a difference. The parity-violating $\rho$-proton interaction constant is
accurately constrained by parity violation in proton--proton scattering,
while the parity-violating pion interaction constant
is constrained by the measurement of parity violation in the reaction $np\rightarrow d\gamma$~\cite{gardner2026hadronic}. A consistent measurement of $g_n$ would confirm the accuracy of future calculations of the $^{229}$Th nuclear structure. 

The intrinsic clock-transition amplitude must include admixtures of \emph{both}
clock states,
\begin{eqnarray}
 D_{\mathrm{PV}}^{\mathrm{intr}}={}&
 \sum_{n,K=5/2^-}
 \frac{\langle g|H_W|n\rangle
    \langle n|D_{+1}|m\rangle}{E_g-E_n}
 \nonumber\\
 &+\sum_{n,K=3/2^-}
 \frac{\langle g|D_{+1}|n\rangle
    \langle n|H_W|m\rangle}{E_m-E_n}.
 \label{eq:pv-e1-sum}
\end{eqnarray}
Removal of the center-of-mass coordinate gives the
valence-neutron E1 effective charge
$e_n^{\mathrm{eff}}=-(Z/A)e=-0.3930e$ and hence
$\mathbf D=e_n^{\mathrm{eff}}\mathbf r$.

For a constant density and spin-independent potential,
$\bm{\sigma}\cdot\mathbf p=i m_N[H,
\bm{\sigma}\cdot\mathbf r]$. Using closure, we obtain for the contribution of the weak mixing between the ground state $\langle g |$ and opposite-parity excited states $|n\rangle$
\begin{eqnarray}
 D_g=\sum_{n}
 \frac{\langle g|H_W|n\rangle
    \langle n|D_{+1}|m\rangle}{E_g-E_n}=
    \nonumber\\
    -i \frac{G_Fg_n\rho_0}{\sqrt{2}} \,\langle g| (\bm{\sigma}\cdot\bm{r}) D_{+1} 
    |m\rangle. 
 \label{eq:Dg}
\end{eqnarray}
A similar calculation of the contribution of the weak mixing between the metastable isomer state $\langle m |$ and opposite-parity excited states $|n\rangle$ gives $D_m=- D_g$. Such cancellation really happens for the diagonal matrix elements, e.g. $D_{gg}$, which give static electric dipole moments forbidden by the time-reversal invariance. However, there is no reason that in the electromagnetic amplitudes between different states there should be exact cancellation between $D_m$ and $D_g$. This cancellation is only a result of the approximate model used. Indeed, the spin-orbit interaction and many-body effects (which we neglected), the spectrum of the opposite-parity states (including nearby states) and electromagnetic amplitudes to these states are significantly different in the ground state and metastable state, which have different spins, 5/2 and 3/2. Therefore, for an order-of-magnitude estimate we may keep one of these contributions only.

Calculations with the oscillator states Eq. (\ref{eq:pv-clock-configurations}) give the following reduced matrix element in the laboratory reference frame: 
\begin{equation}
\langle 5/2|| (\bm{\sigma}\cdot\bm{r})D||3/2\rangle= 11 \,e \, \mathrm{fm}^{2}, 
\end{equation}
and the corresponding parity-violating E1 amplitude 
\begin{equation}
 \langle 5/2|| E1_{\mathrm{PV}} ||3/2\rangle= 5\times10^{-7} i \,e \, g_n \, \mathrm{fm}.
\end{equation}
The corresponding M1 amplitude is~\cite{pvThShigekawa2021}
\begin{equation}
 \langle 5/2|| M1 ||3/2\rangle= 0.237(25) \mu_N=0.0249 (27) \,e \, \mathrm{fm}.
\end{equation}
This gives 
\begin{equation}
 |R|=\left|\frac{E1_{\mathrm{PV}}}{M1}\right|
 =2 \times10^{-5} g_n\,.
 \label{eq:R}
\end{equation}
Separately, we estimated the contribution of the weak mixing between the ground $5/2^+$ state and the closest opposite-parity 146\,keV [752]$5/2^-$ state and found that, despite the relatively small energy denominators in Eq.~(\ref{eq:pv-e1-sum}), this contribution is significantly smaller than the sum of all admixtures to the ground state estimated above. The reason is that the weak mixing matrix element between the ground $5/2^+$ state and $5/2^-$ state and the E1 amplitude from this $5/2^-$ state to the metastable $3/2^+$ state $|m\rangle$ are small (they vanish in the pure oscillator model and require to include the spin-orbit interaction and the $l^2$ term to correct the pure oscillator model). The situation for the weak mixing with the $|m\rangle$ state is similar. 

We remind the reader that Eq. (\ref{eq:R}) is an order of magnitude estimate only. We neglected the contribution of the weak mixing with the $3/2^+$ metastable state, spin-orbit and Coriolis interactions, pairing, quasiparticle--phonon mixing, core polarization and the finite surface of $\rho(r)$. In particular, the actual value of $R$ may be a few times smaller due to a possible partial cancellation between the contributions of the weak mixing in the ground and metastable states.

The here estimated value of parity mixing of the lowest energy states in $^{229}$Th is in a similar range to what has been calculated and measured in transitions of other nuclei such as $^{57}$Fe, $^{175}$Lu, $^{181}$Ta, and $^{203}$Tl~\cite{henley1969parity}. Compared to the value of $|R|=\left|E1_{\mathrm{PV}}/M1\right|$ 
in the transitions between \textit{atomic} levels of Tl, Pb, and Bi  ~\cite{ginges_violations_2004},
$R$ in the nuclear transition of $^{229}$Th, given by Eq. (\ref{eq:R}), is around two orders of magnitude larger.

\section{Observables}
The small parity-violating mixing enables an E1 transition to occur between the ground and isomer state. Its amplitude $|\text{E1}|^2\propto R^2$ is, however, too small to measure directly. Instead, one can use the interference between the E1 and classically allowed M1 transition, which scales linearly with $R$. One possible detection method is to spin-polarize a large ensemble of $^{229}$Th nuclei in their excited state. The subsequently emitted photons during the decay will show a spatial asymmetry that is directly proportional to $R$~\cite{krane1971parity}. A second method is to perform ellipsometry on the fluorescence photons emitted by an unpolarized excited $^{229}$Th ensemble. While pure M1 light would be completely unpolarized, the interference with the E1 transition leads to a small circularly polarized component~\cite{henley1969parity}. Finally, parity violation can cause optical rotation of laser light that traverses a $^{229}$Th ensemble. In the following, we focus on this method as it seems to be the most viable with currently available technologies.
\subsection{Parity-violating nuclear optical rotation }
When light with a wavelength close to the isomer transition frequency traverses a $^{229}$Th-containing sample, the interference between the M1 and parity-violating E1 transitions leads to slightly different complex refractive indices for left- and right-circularly polarized light. This, in turn, causes optical rotation of linearly polarized light by an angle~\cite{zel1959parity, fortson1984atomic} (see Fig. \ref{fig:energy} \textbf{c}):
\begin{equation}
\phi_{\text{PV}}(\Delta)\approx\frac{4\pi l}{\lambda_{Th}}(n(\Delta)-n_0) R\,,
\end{equation}
where $l$ is the length of the sample, $\lambda_{Th}$ is the wavelength of the thorium transition, $\Delta=\nu-\nu_{Th}$ is the frequency detuning of the light, $n(\Delta)$ is the refractive index close to the line and $n_0$ is the refractive index far-detuned from it. For a Lorentzian absorption line with an attenuation coefficient $\alpha$ and half-width at half-maximum (HWHM) $\gamma$, we have $n(\Delta)-n_0=-\frac{\alpha\lambda_{Th}}{4\pi} \frac{\gamma\Delta}{\Delta^2+\gamma^2}$. Between the extrema at $\Delta=\pm\gamma$, the change in optical rotation is:
\begin{equation}
  \Delta\phi_{\text{PV}}=\alpha l R.
\end{equation}
The $^{229}$Th-doped CaF$_2$ crystal X2, which has been characterized in previous works~\cite{Beeks:2022, ooi2026frequency, hiraki2026laser}, shows an attenuation coefficient of $\alpha=1.8\times10^{-3}/$mm at the strongest absorption line~\cite{morawetz2026continuous, Wang2026solid}. A $3\,$cm long fiber with a similar doping density is expected to yield $\alpha l\approx0.05$. Using $R=2\times10^{-5}$, this gives a PVNOR of $\Delta\phi_{\text{PV}}=1\times10^{-6}\,$rad.

\section{Measurement of small optical rotations}
To measure this small PVNOR, we can adapt a method that has been used in atomic vapors to detect optical rotations in the $10^{-7}\,$rad range~\cite{fortson1984atomic}. For this, the  $^{229}$Th-doped crystal is placed between two crossed polarizers. Without optical rotation, the light power leaking through this configuration $P_{\text{L}}$ is given by $P_{\text{L}}=P_0\phi_{\text{E}}^2$, where $P_0$ is the power of the incident light (after absorption losses), and $\phi_{\text{E}}^2$ is the extinction ratio that depends on the quality of the polarizers. PVNOR by an angle $\phi_{\text{PV}}$ increases the leakage to $P_{\text{L}}=P_0(\phi_{\text{E}}^2+\phi_{\text{PV}}^2)$. Practically, it would be very difficult to detect this change in transmitted power, which depends quadratically on the size of the parity violation. To increase the sensitivity of the measurement, one can rotate one of the polarizers slightly by a bias angle $\phi_{\text{B}}$, giving $P_{\text{L}}=P_0(\phi_{\text{E}}^2+(\phi_{\text{PV}}+\phi_{\text{B}})^2)$. That way, $P_{\text{L}}$ varies linearly with $\phi_{\text{PV}}$. For $\phi_{\text{E}}\ll\phi_{\text{B}}\ll1$, the shot-noise-limited signal-to-noise ratio (SNR) is given by $\text{SNR}\approx2\sqrt{\varepsilon P_0\tau/(h\nu_{Th})}\phi_{\text{PV}}$. Here, $\varepsilon$ is the detector efficiency ($0.1$ to $0.2$ for common types of photomultiplier tubes), $\tau$ is the averaging time, $h$ is the Planck constant, and $\nu_{Th}$ is the $^{229}$Th excitation frequency.

Using $P_0=50\,$nW laser, $\phi_{\text{E}}^2=10^{-4}$, and $\phi_{\text{B}}=10^{-1}$ a laser power of $500\,$pW will leak through the polarizers. The PVNOR effect modifies this power by 1\,fW. This is in a range that can be detected with common photomultiplier tubes. For these parameters, and taking $\varepsilon=0.2$, after $10\,$min of data acquisition, a shot-noise-limited SNR of $4$ can be reached.
\begin{figure}[t]
\centering
\includegraphics[width=0.5\textwidth]{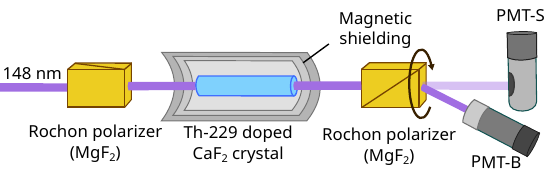}
\caption{Setup for measuring the PVNOR. The laser beam is first polarized. Then it traverses the $^{229}$Th-doped crystal fiber in which it undergoes the PVNOR. A second polarizer is rotated to an angle of $\pi/2+\phi_{\text{B}}$ with respect to the first. The signal that leaks through this second polarizer is measured with a photomultiplier tube (PMT-S). An dditional bias detector (PMT-B) can be used to normalize for fluctuations in the laser power. 
}
\label{fig:setup}
\end{figure}
\section{Systematic effects}
\subsection{Birefringence}
Crystal birefringence leads to a phase retardation between different polarization states. In the general case, this turns the incoming linear polarization into an elliptical polarization state. For the crossed-polarizer detection scheme described above, this poses a systematic effect, because an elliptical polarization appears, at first, indistinguishable from an optical rotation. Strong birefringence can further cause a suppression of the optical rotation. Due to its cubic symmetry, CaF$_2$ does not possess crystallographic birefringence. However, at short wavelengths it possesses spatial‑dispersion‑induced hepta-axial birefringence~\cite{burnett2002symmetry}. To avoid this effect, the CaF$_2$ crystal fibers should be grown and probed along the [111] or [100] directions, which define birefringence-free axes. Any residual crystal birefringence effects can be separated from the PVNOR due to their different frequency dependencies. On the $\sim\!100\,$kHz scale of the $^{229}$Th transition, the magnitude of the crystal birefringence is essentially constant.

This is not the case for a third type of birefringence that is directly related to the $^{229}$Th nuclear transition. The direction of the electric field gradient (EFG) around a $^{229}$Th dopant in relation to the laboratory-frame linear polarization determines how strongly that nucleus absorbs, and therefore also refracts, the incoming light. This defect-induced birefringence follows the same refractive line shape as the PVNOR and could therefore cause a dangerous systematic. Defect-induced birefringence vanishes in crystals containing multiple guest-sites with isotropically oriented EFGs. This is generally the case for defects in CaF$_2$, which occur in orientational variants related by $O_h$ symmetry operations.~\cite{hiraki2026laser, rehn2026defect}.

\subsection{Faraday effect}
The interaction with external magnetic fields can also cause optical rotation at the $^{229}$Th nuclear transition frequency through the Faraday effect. One can distinguish between three distinct contributions~\cite{serber1932theory}. The Faraday $a$-term results from the magnetic-field-induced line splitting and causes optical rotation that is symmetric around the transition frequency (see Fig. \ref{fig:energy} {\bf d}). This facilitates distinguishing it from the PVNOR.

The peak rotation through the $a$-term is $\phi_{F_a}/B\approx\alpha l\mu_N/ (h\nu_{Th})\approx4$\,rad T$^{-1}$, where $B$ is the magnetic field strength and $\mu_N$ is the nuclear magneton. To suppress this effect below the strength of the PVNOR, the sample should be magnetically shielded to $B<100\,$nT, which is routinely achieved. The Faraday $b$-term is caused by the magnetic-field-dependent mixing with nearby states that modifies the transition probabilities. In contrast to the $a$-term, it is asymmetric and therefore harder to distinguish from the PVNOR. The relative strength between $a$-, and $b$-term can be estimated through $\phi_{F_b}/\phi_{F_a}\sim\Gamma/\Delta\omega$, where $\Delta\omega$ is the splitting between the different $|m|$ states~\cite{papchenko1994calculation}. 

In $^{229}$Th-doped CaF$_2$, $\Delta\omega\sim100$ MHz. Consequently, $\phi_{F_b}/\phi_{F_a}\sim10^{-3}$, meaning that the asymmetric effect will be sufficiently suppressed if magnetic shielding is used. 

Finally, the Faraday $c$-term originates from inhomogeneous equilibrium state populations. It as well is asymmetric. Its strength can be estimated through $\phi_{F_c}/\phi_{F_a}\sim \hbar\Gamma /k_B T$, where $k_B$ is the Boltzmann constant, and $T$ is the crystal temperature~\cite{papchenko1994calculation}. At room temperature, $\phi_{F_c}/\phi_{F_a}\sim 10^{-8}$. The $c$-term is therefore negligible for our purposes.

\subsection{Perturbation through defects}
In previous searches for fundamental symmetry violations using crystals, local breaking of the crystal symmetry was a major concern~\cite{bialek1986macroscopic, lamoreaux2002solid, heidenreich2005limit, eckel2012limit, nima2026limit}. As with birefringence, spurious crystal effects can be well separated from the true signal due to its unique frequency response. Generally, more dangerous are crystal defects that break local inversion symmetry and perturb the $^{229}$Th nuclear wavefunctions directly. The PV Hamiltonian $H_{\text{PV}}$ mixes states with a magnitude of $\sim 1\,$eV. Interactions with crystal fields are generally much lower. E.g., the quadrupole shift in CaF$_2$ is below $1\,\mu$eV. Every external field interaction that mimics PV enters the perturbation calculation through expressions analogous to Eq. (\ref{eq:pv-e1-sum}). Therefore, the PV state mixing in nuclei is very robust against perturbations through external fields such as those from crystal defects. This property is quite different from atoms or molecules for which the state mixing in low external electric fields usually surpasses the strength of internal parity violations.

\section{Discussion}
The sensitivity of the measurement could be increased by several means. In other host crystals in which thorium is part of the stoichiometric crystal structure, such as ThF$_4$, one can expect a strongly reduced line broadening and therefore more strongly peaked absorption lines. With these, sub-cm crystals with an optical depth of $\alpha l\sim1$ could be grown, increasing the PVNOR by a factor of 20. However, in such non-cubic solids, birefringence effects are more difficult to avoid. Increasing the laser power decreases the required measurement time linearly. With a laser power of $1\,\mu$W \cite{huang2026nuclear} and $\alpha l=1$, an SNR of $4$ for a peak-to-peak measurement could already be reached after less than $100\,$ms.

\textit{Acknowledgments.} -- Work of VF was supported by the Australian Research Council Grant No.\ DP230101058. Part of this work has been funded by the European Research Council (ERC) under the European Union’s Horizon 2020 research and innovation programme (Grant Agreement No. 856415 and 101266822) and the Austrian Science Fund (FWF) [Grant DOI: 10.55776/KIN4315625 and  10.55776/F1004]. The project 23FUN03 HIOC [Grant DOI: 10.13039/100019599] has received funding from the European Partnership on Metrology, co-financed from the European Union’s Horizon Europe Research and Innovation Program and by the Participating States.

\bibliography{biblio}

\end{document}